\documentclass[aps,pra,twocolumn,floatfix,floats,showpacs,superscriptaddress,raggedbottom]{revtex4}
\usepackage{amsmath,amsfonts,amssymb,amsthm}
\usepackage{color,ulem}
\usepackage{graphicx,latexsym}
\pdfoutput=1
\begin{document}

\title{Nonlinear interference in a mean-field quantum model}

\author{Gilbert Reinisch}
\affiliation{Universit\'e de Nice - Sophia Antipolis, CNRS,\\ Observatoire de la C\^ote d'Azur,
             BP 4229, 06304 - Nice C\'edex 4, France}
\author{Vidar Gudmundsson}
\affiliation{Science Institute, University of Iceland, Dunhaga 3, IS-107 Reykjavik,
             Iceland}

\begin{abstract}
      Using similar nonlinear stationary mean-field models
      for Bose-Einstein Condensation of cold
      atoms and interacting electrons in a Quantum Dot, we propose to describe the
      original many-particle ground state as a one-particle statistical mixed state
      of the nonlinear eigenstates whose weights are provided by the eigenstate non-orthogonality.
      We search for physical grounds in the interpretation
      of our two main results, namely, quantum-classical nonlinear
      transition and interference between nonlinear eigenstates.
\end{abstract}


\maketitle

\section{Introduction}
Commonly, quantum mechanical models of many interacting
particles --
linear models -- are made computationally tractable by transforming them
into a mean-field single-particle models, at the cost of making the model
nonlinear, but solvable \cite{Kohn98:1253}.
Yet, solvable has a different meaning within two fields of condensed matter physics,
the field of cold atoms in an external trap, and the field
of electrons in a quantum dot. In the former field, the ensuing equation for
the Bose-Einstein condensate (BEC), namely, the Gross-Pitaevskii equation (GPE),
is often
solved directly as a nonlinear differential equation
\cite{Dalfovo99:463,Leggett01:307}. In the latter case, for
electrons confined in a quantum dot, the approximation applied for the
Coulomb interaction
between the electrons leads to the Hartree, the Hartree-Fock, or Kohn-Sham
type of an equation. All these are nonlinear equations  that are conventionally
solved by an iteration scheme. These equations for the electrons have a
similar nonlinear term as for the GPE, but the long range nature of the
Coulomb interaction makes them more complex. Within each iteration, the
mean-field potential is constructed with information about the charge
distribution and/or the wave functions from former iterations.
The nonlinear Schr{\"o}dinger-type equation is thus considered as a linear
equation within each iteration step and solved by means of methods from
linear algebra supplying a orthonormal set of eigenfunctions and
corresponding eigenvalues. The nonlinear behavior enters the procedure once again when
the eigenfunctions are used to calculate for the next step the new mean-field potential the
electron is moving in. After the fulfillment of some criteria of convergence
for the iterations, the end product is a solution in terms of
orthonormal wave functions, presumably
representing the single-electron states of the problem.

In applications in quantum chemistry, it is often more appropriate to resort to
functional basis that are not orthogonal in order to
reduce the size of the numerical effort. A direct solution of the
equations then leads to a general eigenvalue problem and solutions that are
not orthonormal, but usually the orthonormality of the solutions is restored
by a refined handling of the general eigenvalue problem \cite{Luana10:101}.

For the fermion system (e.g.\ the electrons in a quantum dot), the hope is that
the mean-field single-electron solutions that have been achieved by a self-consistent
iteration of the Schr{\"o}dinger equation and the equation for the mean-field potential
reflects in some  reasonable approximation properties of the original huge
linear many-electron problem. Of course the individual single-electron states,
wave functions, or orbitals are of limited value, but they can be used to
construct more physically relevant entities like various response functions,
the total energy, and the total charge
distribution. Confidence in the method has come from comparison to
numerical solutions of the corresponding truncated many-electron problem for
few electrons \cite{Maksym90:108,Pfannkuche93:2244}, and experimental results.

For the boson system, namely, cold atoms in a trap, the ground state of the
nonlinear GPE has been calculated in order to gain information about properties
of the BEC, and the GPE has been generalized to finite temperatures by the means
of self-consistent Hartree-Fock or Bogolyubov approximations where special care
has been taken to construct orthonormal states for the system
\cite{Huse82:137}; otherwise,
reasons are given for neglecting the fact that a particular approximation
does not preserve the orthogonality \cite{Goldman81:2870}.

After utilizing these methods for years in each of the mentioned subfields and
comparing the different approaches, we would like to draw attention of the reader
to interesting open questions concerning the nonlinearity of the underlying
equations.  In particular, we show that the nonlinear
mean-field single-particle ground state
is \underbar{not} a pure state where all boson particles would condense,
like in standard linear quantum theory. It is a
\underbar{mixed state} that allows a small part (less than one percent) of the
boson gas to populate higher nonlinear excited energy levels.
We present a simple model to highlight our concerns.
We are of course aware that these mean-field equations do only asymptotically
describe the respective systems within an appropriate range of physical
parameters like many other celebrated equations in physics, even though
they have been found to reproduce the properties of the
systems outside of the parameter range that the approximation itself can
be justified for.

In the summary section we reflect on the physical relevance and interpretation
of our findings.

\maketitle

\section{The nonlinear Schr{\"o}dinger quantum model}
Prior to the discovery of the microscopic BCS theory of superconductivity
as a Bose condensation of electrons paired by microscopic electron-lattice
interactions \cite{Bardeen57:162}, tentative links of
superconductivity with the interaction between electrons and lattice vibrations had
already been explored \cite{Froehlich50:845} \cite{Bardeen51:261}.
Most interesting in the scope of the present work is Schafroth's early suggestion that
charge-carrying bosons in a metal at low temperature  actually constitute a gas of
bound two-electron states \cite{Schafroth54:1442,Schafroth54:1149} which can then be
described by a self-consistent normalized nonlinear Schr{\"o}dinger-Poisson
(SP) differential system \cite{Schafroth55:463}. At that time (1955),
Schafroth just wanted to emphasize the role of long-range Coulomb interactions between
bosons by use of mostly qualitative arguments. He assumed this charged boson  gas to
be defined by the
stationary Schr{\"o}dinger equation and proposed a one-dimensional (1D) SP
model which  considers the particles as moving in an effective potential ${\cal V}(x)$
related by Poisson's equation to the local boson charge density with an additional
uniform source term modeling
both the charge density of the background (in order to make the whole system
electrically neutral) and the non-condensed particles. Evidently, this
uniform source term in 1D Poisson equation can be regarded as
the second derivative of an additional external harmonic confining potential.
Consequently, this 1954 SP  model was probably the first
ab-initio mean-field nonlinear self-consistent attempt to describe the competition that occurs
in a charged boson gas between external parabolic particle
confinement and long-range internal particle-particle Coulomb repulsion.

Even in 1D, the numerical solution of the SP differential system was not available at
that time. Hence, the author considered the classical Thomas-Fermi (ThF) approximation
where both the effective potential ${\cal V}(x)$ and the boson gas wave function
$\Psi$ were assumed uniform and respectively equal to the energy shift of the ground
state due to Coulomb repulsion and to the wave function's averaged value $L^{-1/2}$
over large regions of spatial extension $L$. These results led to the
early conclusion that the occurrence of superconductivity in metals is indeed
due to the formation of some kind of resonant two-electron boson states
at low temperature;  a conjecture which soon became validated by the superconductivity
BCS model as well as by Bogoliubov-type investigations of a Coulomb Bose gas
\cite{Foldy61:649,Alexandrov95:5887}.

The present work revisits, by use of  straightforward numerical routines,
this pioneering SP differential model of a charged boson gas of particle mass $M$
trapped in the radial axisymmetric parabolic potential $V({\bf x})\equiv V(r)= {1\over 2}
M\omega^2r^2 $. As the simplest such example, let us mention opposite-spin Cooper-like
electron pairs in quantum-dot helium \cite{Pfannkuche93:2244}.
Inasmuch as we wish to emphasize interference-like quantum coherence effects due to
nonlinearity, we extend our investigation to the neutral hard-sphere GPE boson gas.
Indeed, the time-dependent interference between two BEC's has been investigated in
the one-dimensional case by exact solutions of the GPE obtained with methods of
inverse scattering,
thus displaying interference as an actual nonlinear interaction
between GPE envelope solitons \cite{Liu00:2294}.
In this work, we also consider the complementary model of a single,
harmonically-trapped, stationary BEC with discrete non-orthogonal
(since the equation is nonlinear) interfering eigenstates $|\Psi_i\rangle$.
Due to 2D radial axisymmetry, $|\Psi_i\rangle = |\psi_i\rangle\otimes|m\rangle$
is an eigenstate of the angular-momentum operator
with  the eigenvalue $m\hbar$. As $\langle\Psi_i|\Psi_{j}\rangle =
\langle\psi_i|\psi_{j}\rangle\langle m_i|m_{j}\rangle\neq 0$ only if
$m_i=m_j=0$, it is sufficient to consider the zero-angular-momentum states
$m=0$ in order to
display peculiar physical properties related to Bose eigenstate non-orthogonality
(in contrast with, e.g., BEC vortex-nucleation where the $m=1$ nonlinear
eigenstate, actually orthogonal to the $m=0$ ground state, plays the major role
\cite{Reinisch07:120402}). The discrete radial one-particle Bose eigenfunction
$\psi_i(r)$ is defined in the mean-field approximation by the
stationary Schr{\"o}dinger equation related to the corresponding chemical-potential
eigenvalue $\mu_i$. In scaled form $u_i(r)\propto \psi_i(r)$, it reads:
\begin{equation}
\label{eq-Schroe}
      {\ddot u_i}+{1\over X}{\dot u_i} +
      \Bigl[{\tilde \mu_i}-{\tilde \Phi_i}-{X^2\over 4} \Bigr]\, u_i=0,
\end{equation}
where the dot stands for derivation with respect to the radius $X$ measured in units
of the characteristic length $l_0=\sqrt{\hbar/2 M\omega}$ and the tilde
superscript labels energy in units of $\hbar\omega$. The dimensionless
particle-particle interaction energy
${\tilde \Phi}_i(r)$ per particle introduces into the system
nonlinearity which is appropriately scaled to unity. Indeed, defining
\begin{equation}
\label{eq-defOFu}
      u_i(r)=\sqrt{\frac{\pi\hbar{\cal N}_i}{M\omega}}\psi_i(r),
\end{equation}
where the dimensionless parameter ${\cal N}_i$ is self-consistently given
by the solution of Eq.\ (\ref{eq-Schroe}) according to
\begin{equation}
\label{eq-normOFu}
      \int_0^{\infty}u_i^2\, XdX= {\cal N}_i,
\end{equation}
we obtain the necessary 2D axisymmetric normalization condition
\begin{equation}
\label{eq-normOFpsi}
      \int\,|\Psi_i|^2\,d^2{\bf x}=\int\,\psi_i^2\,
      2\pi r dr=1,
\end{equation}
together with the scaling to unity of the nonlinear coefficient for both the GPE:
\begin{equation}
\label{eq-nlGPE}
      {\tilde \Phi}_i=u_i^2,
\end{equation}
and the SP differential system
\begin{equation}
\label{eq-nlSP}
      -{\ddot {\tilde \Phi}_i}-{1\over X}{\dot {\tilde \Phi}_i} =u_i^2.
\end{equation}
Equation (\ref{eq-nlGPE}) defines the contact interactions of the hard-sphere BEC while the 2D
Poisson Eq.\ (\ref{eq-nlSP}) defines the long-range Coulomb interactions of the
charged Bose gas (e.g.\ quantum-dot helium \cite{Pfannkuche93:2244}). Practically,
the value of the dimensionless norm (\ref{eq-normOFu}) is given
by the experimental conditions. We have ${\cal N}_i= 4N(a_s/L_z)$ for the BEC defined by
its  scattering length $a_s$ and extension $L_z$ in the axial $z$-direction that contains $N$ particles
in the $i$th nonlinear eigenstate \cite{Reinisch07:120402}. Specifically,
${\cal N}_1= 30.29$ for the ground state in  Paris-ENS' large laser-beam stirrer experiment
\cite{Madison00:806,Chevy00:2223,Madison01:4443} while
${\cal N}_0= 187.35$ in Boulder-JILA's rotating normal-cloud experiment
\cite{Matthews99:2498,Haljan01:210403}. These high values  ${\cal N}_1\gg1$   of
the nonlinear parameter
are due to the number of atoms in the trap which is quite large (typically of order $10^{4-5}$).
They allow the ground state to be approximated by its
localized ``negative-curvature'' parabolic Thomas-Fermi profile
\cite{Dalfovo99:463,Leggett01:307} and define in the present work the so-called
``classical-nonlinear'' regime. On the other hand, the experimental parameters
defining quantum-dot helium yields  much lower values ${\cal N}_1\sim2\,-\, 3$ for the
ground state nonlinearity \cite{Reinisch09:4650}; they would correspond in our study to
the ``quantum-nonlinear'' regime where interference effects occur.
\section{Discussion}
\begin{figure}[htbq]
      \includegraphics[width=0.38\textwidth,angle=0,viewport=120 110 2000 1700,clip]{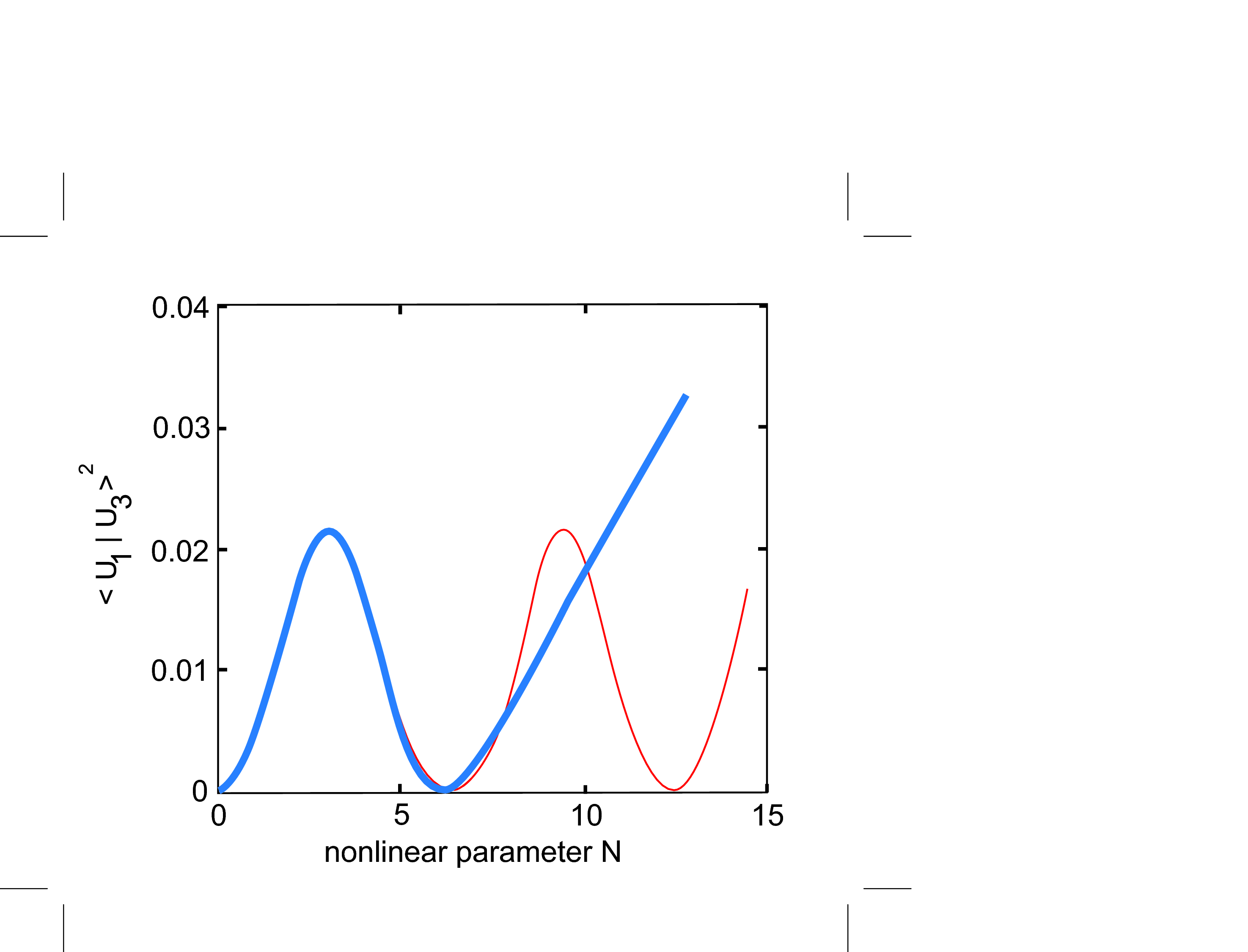}
      \caption{The quantum-dot helium interference pattern defined by the square probability
               amplitude $w^{13}=\langle u^1|u^3\rangle^2$, as compared with its
               $\sin^2$ approximation defined by Eq.\ (\ref{eq-2Dinterf}) (thin continuous line).
               There occurs a clearly visible threshold at ${\cal N}\sim 6$ where
               the system bifurcates from its interfering
               quantum regime towards its Thomas-Fermi classical one.}
\label{fig1}
\end{figure}

The interest in nonlinear quantum eigenstates can be illustrated for both
the charged Bose gas defined by Eq.\ (\ref{eq-nlSP}) and the GPE system
defined  by Eq.\ (\ref{eq-nlGPE}). In the former case,
the nonlinear eigenstates can be regarded as unperturbed although they take
into account, in addition to the particles' external parabolic confinement,
the (usually quite important) long-range
Coulomb interactions \cite{Cohen88:04}.
In the later case, nonlinear quantum eigenstates
in the physical description of very-many-particle stationary
BEC's are  unavoidable. Indeed, there is
practically no other choice than solving the nonlinear GPE
(\ref{eq-Schroe}) and (\ref{eq-nlGPE}).
Therefore we numerically consider the discrete real-valued
radial-symmetric normalized eigenstates $u_i$ of Eqs.\ (\ref{eq-Schroe}) and
(\ref{eq-nlGPE})-(\ref{eq-nlSP}) which are by choice
non-orthogonal since we restrict ourselves to
$m_i=0$ zero-angular-momentum $s$ states \cite{Reinisch04:033613}.
Specifically, we consider the two first such  eigenstates
$|u^1\rangle$ and $|u^3\rangle$ whose superscript
labels refer to their number of $\hbar\omega$
eigenquanta in the limit of vanishing nonlinearity, i.e.\ for the 2D linear parabolic
system \cite{Pfannkuche93:2244}. Their normalized inner product
$\langle u^1 | u^3\rangle\neq 0$ yields the
statistical weight  $w^{13}=w^{31}=| \langle u^1 | u^3\rangle|^2 =
\langle u^1 | u^3\rangle^2\neq 0$ that defines the
\underbar{mixed} ground state whose appropriate
description in terms of the nonlinear density matrix $\rho$ is
\begin{equation}
\label{eq-mixed}
      \rho=\frac{1}{1+w^{13}}|u^1\rangle \langle u^1| +
      \frac{w^{13}}{1+w^{13}}|u^3\rangle \langle u^3| .
\end{equation}
As a matter of fact, the standard use of a density matrix demands an orthonormal basis
of eigenstates. Therefore, Eq.\ (\ref{eq-mixed}) should be understood as an extension
of this concept to the nonlinear case where the eigenstates are not orthogonal. This generalization
is all the more natural as the deviation from orthogonality is small (see below Fig.\ \ref{fig1})
and it yields a transparent physical interpretation. Indeed, Eq.\ (\ref{eq-mixed})
suggests that the probability per particle for the Bose gas to
be in the nonlinear excited state $|u^3\rangle$ is but its probability
$ 1/ (1+w^{13})$ to be in the nonlinear ground state $|u^1\rangle$ multiplied by the
mere transition probability $w^{13}$ from $|u^1\rangle$
to $|u^3\rangle$. However this last expression must be taken
in a somewhat loose sense since the present time-independent description is, strictly
speaking, incompatible with the concept of a quantum transition (we will return
to this discussion further below).

The paradoxical property, compared with standard linear theory, is that as the
system is in the mixed state described by Eq.\ (\ref{eq-mixed}), it nevertheless
allows interference between the two eigenstates
$|u^1\rangle$ and $|u^3\rangle$ to occur, according to the following theorem:
\begin{equation}
\label{eq-theor}
      \langle u^1 | u^3\rangle=\frac{\Phi^1_{31}}{\mu_1-\mu_3} +
                               \frac{\Phi^3_{13}}{\mu_3-\mu_1}.
\end{equation}
The subscripts in $\Phi^i_{jk}$ define the
Hilbertian matrix elements of the particle-particle interaction potential
$\Phi^{i}$  corresponding to $|u^i\rangle$ (see Eq.\ (\ref{eq-Schroe})).
These elements have been calculated by use of the nonlinear eigenfunctions
$u^i$ while $\mu_i$ are those nonlinear eigenvalues (or equivalently chemical
potentials) which respectively define the $u^i$'s.
Equation (\ref{eq-theor}) is a direct consequence of the Hermiticity
of the Laplacian operator in the Schr\"odinger equation (\ref{eq-Schroe})
\cite{Bec10:00}. The first term on the r.h.s.\ of Eq.\ (\ref{eq-theor}) defines
the probability amplitude for the system in the nonlinear
eigenstate $|u^1\rangle$ to be also in the nonlinear eigenstate
$|u^3\rangle$, due to the interaction potential
$\Phi^1$ defined by the probability density $|u^1|^2=(u^1)^2$ through
Eqs.\ (\ref{eq-nlGPE}) or (\ref{eq-nlSP}), while the second term defines the
reverse process, namely, the probability amplitude for the system in
$|u^3\rangle$ to be also in $|u^1\rangle$ as a consequence of the interaction
potential $\Phi^3$ defined by $(u^3)^2$. Equation (\ref{eq-theor}) is \underbar{exact}.
Therefore no perturbative-like ordering in $\Phi^{i}$  is needed although, of course,
a straightforward time-independent perturbation scheme that considers $\Phi^{i}$ as
perturbation in its respective Schr\"odinger equation
(\ref{eq-Schroe}) allows one to recover it (for instance when nonlinearity is weak).
This point will be further developed in the next part.

The two amplitudes in Eq.\ (\ref{eq-theor}) interfere because the
corresponding $1\leftrightarrow 3$ processes are indistinguishable in the build-up of the
probability amplitude $\langle u^1 |u^3 \rangle$. Therefore this amplitude actually
defines the \underbar{nonlinear quantum coherence} in our two-state system.
The interference pattern is increased by the $\omega\rightarrow 0$
progressive flattening of the parabolic trap in the case of
the confined charged Bose gas and by the increase of the particle number in the GPE system;
or, equivalently for both of them, by a progressive increase of the nonlinearity parameter
${\cal N}$ in the system, as defined by Eq.\ (\ref{eq-normOFu}).

Figure \ref{fig1} illustrates in the case of 2D axisymmetric quantum dot helium
the remarkable behavior of the normalized
mixed-state statistical weight
$w^{13}=w^{31}=| \langle u^1 | u^3\rangle|^2 = \langle u^1 | u^3\rangle^2$
where
\begin{equation}
\label{eq-2Dscalprod}
      \langle u^1|u^3\rangle ={1\over {\cal N}}\,
      \int_0^{\infty}u^1u^3\,XdX.
\end{equation}
The two eigenstates $u^{1,\,3}$ must obviously correspond to the
same physical system. For the GPE defined by Eq.\ (\ref{eq-nlGPE}),
this means that their particle number
should be the same and hence ${\cal N}={\cal N}_1={\cal N}_3$.
The case of the confined charged Bose gas defined by Eq.\ (\ref{eq-nlSP}) demands
an additional condition which states that the external trap
parabolicity $\omega$ is identical for both eigenstates $u^i$.
Nevertheless it also yields ${\cal N}=
{\cal N}_1\sim {\cal N}_3$ \cite{Reinisch09:4650}.
We numerically obtain the simple interference pattern
\begin{equation}
\label{eq-2Dinterf}
         w^{13} = \gamma\,\sin^2 \Biggl({{\cal N} \over 2}\Biggr)
         + o(\gamma^{2}),
\end{equation}
for ${\cal N} \leq 6$ and  $\gamma=2.29\,\,10^{-2}$.
It displays a remarkable nonlinear resonance
where the quantum coherence maximum is reached about the particular value
${\cal N}={\cal N}^*\sim 3$. For quantum-dot helium, the only free
physical parameter is the trap parabolicity $\omega$.
Therefore we have ${\cal N}\equiv {\cal N}(\omega)$
and the resonance occurs for the particular parabolic trap profile
$V(r)={1\over 2}M\omega^* r^2$ where $\omega^*=\omega({\cal N}^*)$.
Numerical simulations in this 2D axisymmetric model show that
$\hbar\omega^* \sim 0.14\,\epsilon$ where $\epsilon=Me^4/\hbar^2$ is the
effective atomic energy unit \cite{Reinisch09:4650}.
Hence $\hbar\omega^* \sim 3.80$ eV for electrons in vacuum since
$\epsilon =27.21$ eV while $\hbar\omega^* \sim 1.66$ meV in
the case of GaAs quantum-dot helium where $\epsilon =  11.86$ meV.

In the case of long-range Coulomb particle-particle interactions described
by Eq.\ (\ref{eq-nlSP}), there exists a maximum amplitude $u_0$ for
both nonlinear eigenstates $u^1$ and $u^3$ that is
clearly visible on Fig.\ 2 when ${\cal N}$ varies from $10^{-2}$ ($u_0\sim 0.1$)
to $10^{2}$ ($u_0\sim 1$). For such high values of ${\cal N}$,
one reaches the classical asymptotic ThF regime defined by neglecting the
quantum kinetic Laplacian derivative terms in Eq.\ (\ref{eq-Schroe}) \cite{Schafroth55:463}.
Then ${\tilde \Phi^i(X)}\sim {\tilde \mu}_i -X^2/4$ and Eq.\ (\ref{eq-nlSP})
yield the common limit $u^1(X)\equiv u^3(X)\equiv u_{TF}\equiv 1$.
\begin{figure}[htbq]
      \includegraphics[width=0.48\textwidth,angle=0,viewport=20 20 1740 1300,clip]{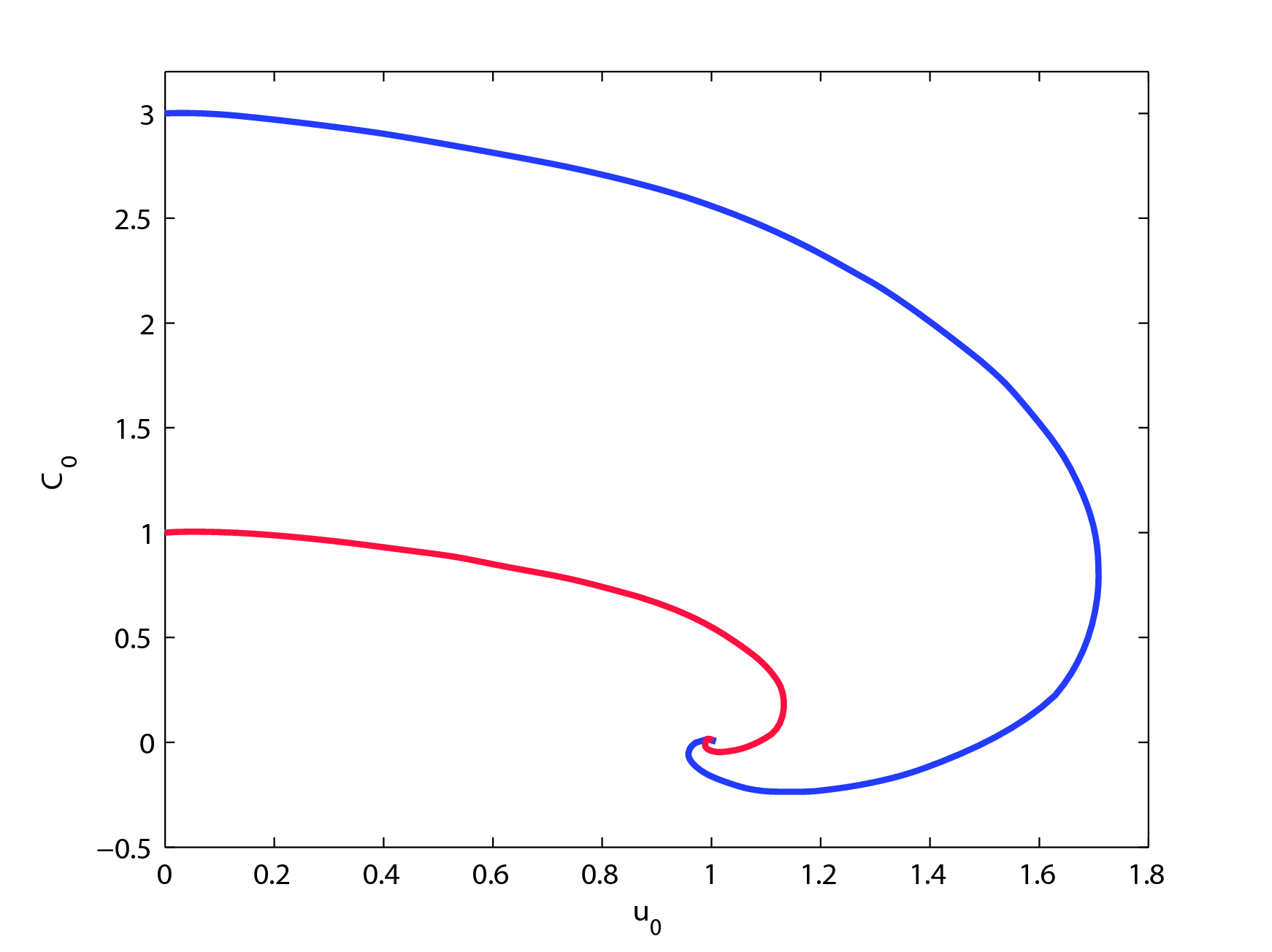}
      \caption{The convergence in the \{$C_0={\tilde \mu}-{\tilde \Phi(0)}$\}
               vs \{$u_0=u(0)$\} initial-condition phase space of the two-level
               SP nonlinear system defined by $u^1$ (lower red curve) and $u^3$
               (upper blue curve) towards
               the common fixed point $u_0=1$ and $C_0=0$ when the nonlinearity parameter
               increases from the linear regime
               ${\cal N}\sim 10^{-2}$ to the classical Thomas-Fermi one defined by
               ${\cal N}\sim 10^2$. The corresponding common asymptotic
               eigenstate profile $u_{TF}(X)$ for these two
               first $m=0$ nonlinear eigenmodes is the uniform
one $u_{TF} \equiv 1$.}
\label{fig2}
\end{figure}
The respective initial conditions of the two modes spiral down
towards the same Thomas-Fermi fixed point for ${\cal N}\rightarrow\infty$. At
the same time, the corresponding eigenstate profiles $u_1$ and $u_3$  start
increasing their width instead of their amplitude: see Fig.\ \ref{fig3}.
There is a progressive merging of $u_1$ and $u_3$ into the single asymptotic
uniform  mode $u_{TF}$ defined by the fixed point
$\{1,\,0 \}$ in Fig.\ \ref{fig2}. Quantum discreteness disappears for
${\cal N}\gg 1$, which is the hallmark of the classical regime.
\begin{figure}[htbq]
      \includegraphics[width=0.48\textwidth,angle=0,viewport=20 20 1740 1250,clip]{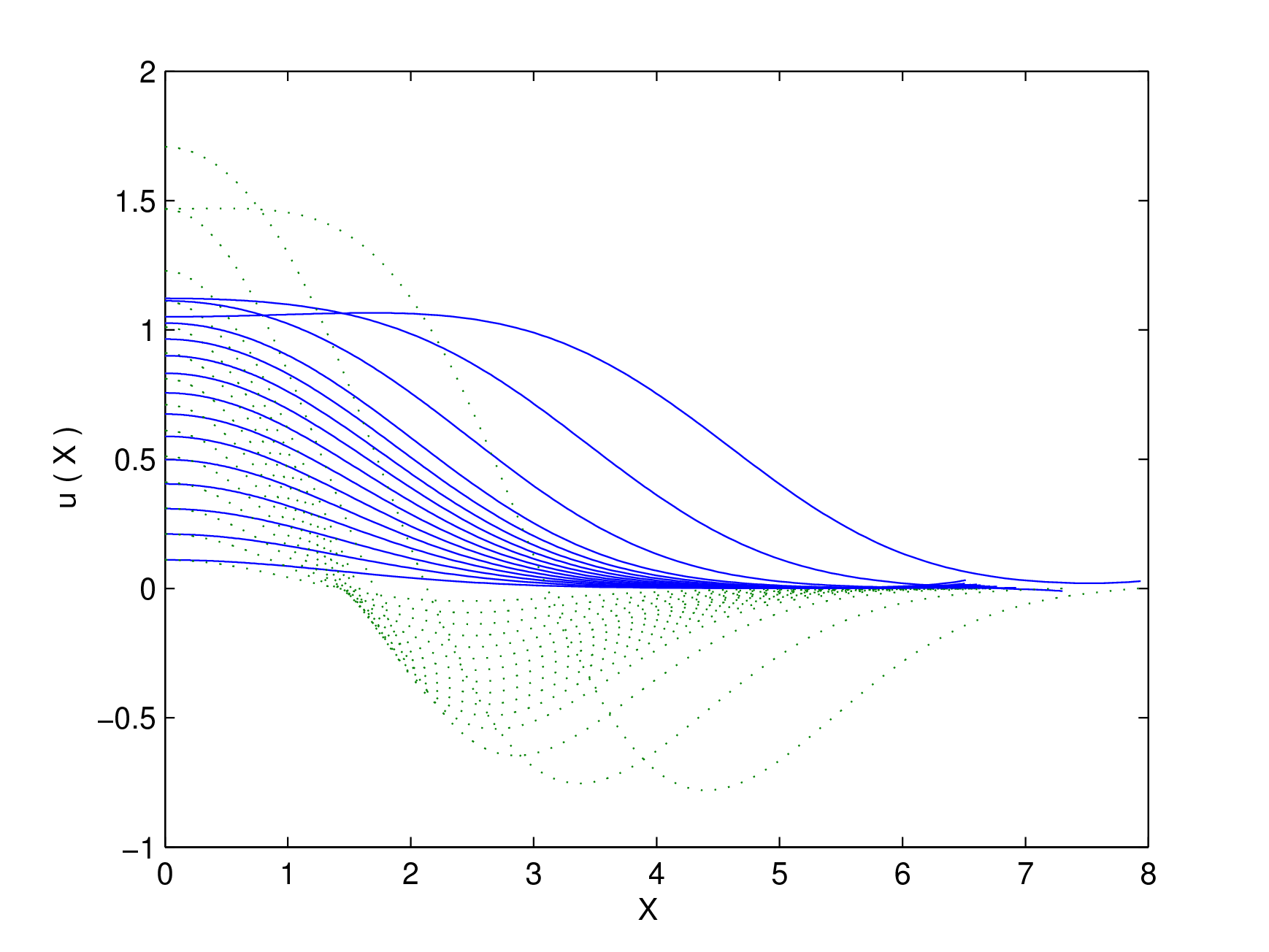}
      \caption{The nonlinear SP eigenstate profiles $u^1$  (continuous)
               and $u^3$ (dashed) for increasing values ${\cal N}\leq 8$ of the
               nonlinearity. Note the last $u^1$ and $u^3$ profiles where
               the width ---instead of the amplitude---  starts increasing.}
\label{fig3}
\end{figure}

\begin{figure}[htbq]
      \includegraphics[width=0.48\textwidth,angle=0]{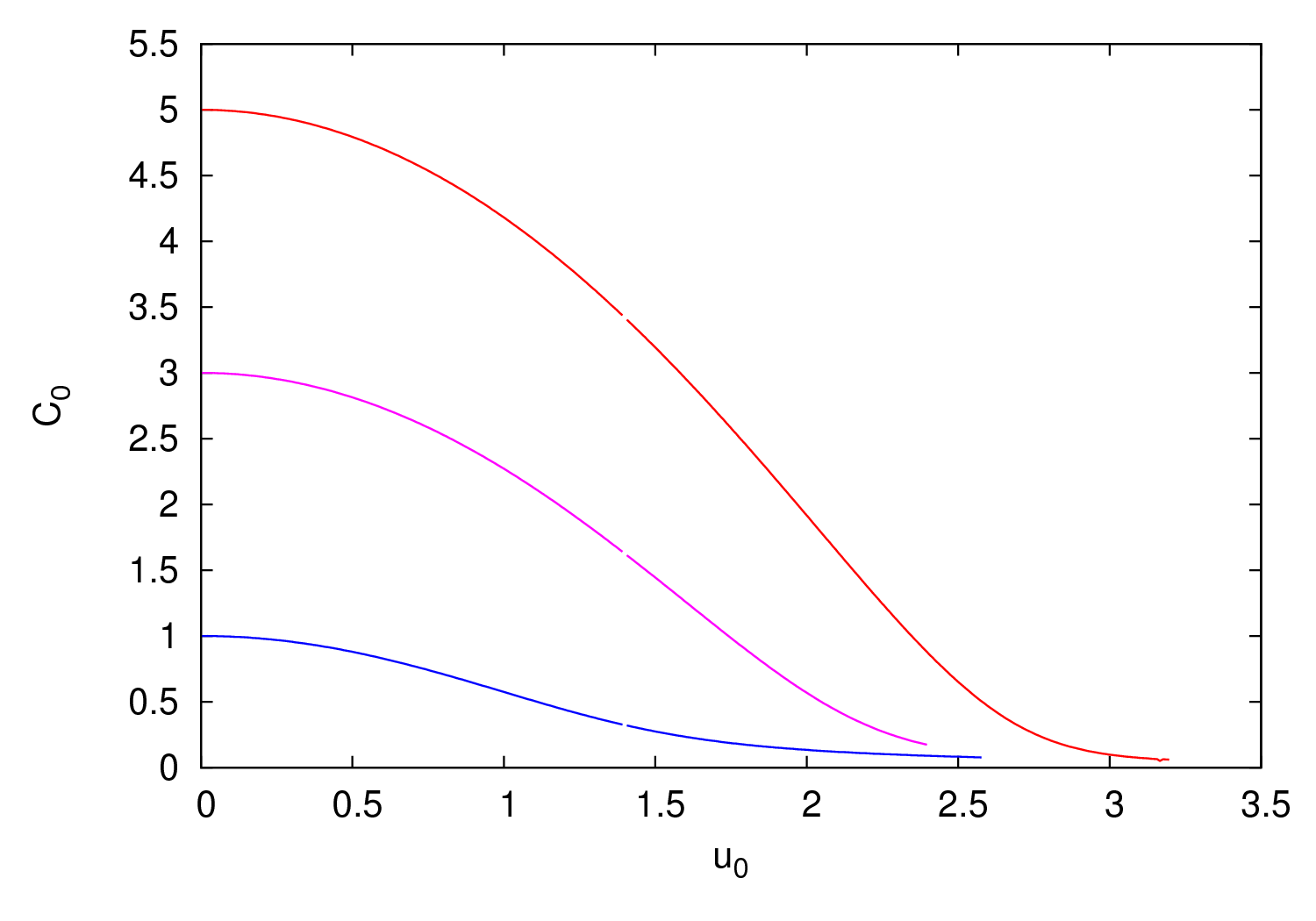}
      \caption{The convergence in the \{$C_0={\tilde \mu}-{\tilde \Phi(0)}={\tilde \mu}-u_0^2$\}
               vs \{$u_0=u(0)$\} initial-condition phase space of the three-level
               GPE system defined by its $m=0$ nonlinear states  $u^1$ (lower blue curve),
               $u^3$ (middle pink curve) and $u^5$ (upper red curve) towards
               the common fixed point $u_0=\infty$ and $C_0=0$ when the nonlinearity parameter
               increases from the linear regime ${\cal N}\sim 10^{-2}$ towards the classical Thomas-Fermi one
               ${\cal N}\gg 1$. The corresponding common asymptotic
               eigenstate profile $u_{TF}(X)$ for these
               nonlinear eigenmodes is defined by $u_{TF} \equiv \sqrt{{\tilde\mu} - X^2/4}$ as
               shown by Eqs.\ \ref{eq-Schroe} and \ref{eq-nlGPE} when the Laplacian derivative
               terms are neglected.}
\label{fig4}
\end{figure}

\begin{figure}[htbq]
      \includegraphics[width=0.48\textwidth,angle=0]{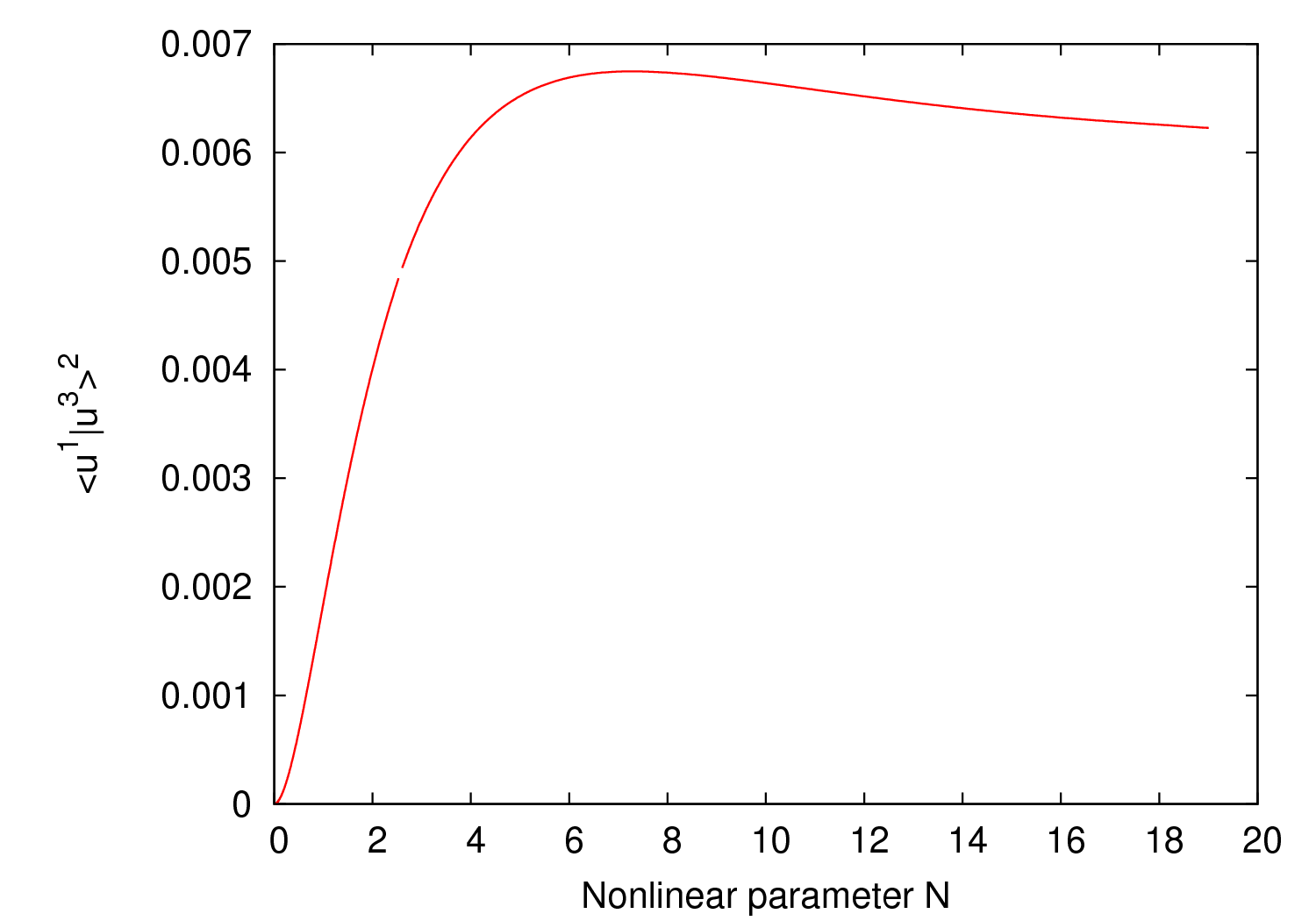}
      \caption{The interference pattern defined by the square probability
               amplitude  $w^{13}=\langle u^1|u^3\rangle^2$ in the case of the GPE system.
               The threshold for the quantum-classical transition occurs
at ${\cal N}\sim 10$.}
\label{fig5}
\end{figure}

Quite different is the convergence of the GPE
system (\ref{eq-nlGPE}) towards the ThF classical regime still defined by
${\tilde \Phi^i(X)}\sim {\tilde \mu_i}-X^2/4$, as shown by
Fig.\ \ref{fig4}. Then we indeed have $\lim_{{\cal N}\rightarrow\infty}C_0=
\lim_{{\cal N}\rightarrow\infty}[{\tilde\mu}-{\tilde\Phi}(0)]=
\lim_{{\cal N}\rightarrow\infty}[{\tilde\mu}-u_0^2]=0$, but the nonlinear
eigenstate amplitudes $u_0$ grow without limit while their corresponding profiles
converge towards $u_{TF}(X) \equiv \sqrt{{\tilde\mu} - X^2/4}$, as
shown by Eqs.\ \ref{eq-Schroe} and \ref{eq-nlGPE} when the Laplacian derivative
terms are neglected. There is no bifurcation from the
amplitude-growing regime to the width-growing one, like in the SP case
displayed by Figs.\ \ref{fig2} and \ref{fig3}.
It can be said that, due to GPE particle-particle contact interactions defined by
Eq.\ (\ref{eq-nlGPE}), the bosons pile up in the trap rather than spread out.
The corresponding nonlinear resonance defined by Eq.\ (\ref{eq-2Dscalprod})
with  ${\cal N}= {\cal N}_1={\cal N}_3$ looks also quite differently, as shown by
Fig.\ \ref{fig5}.
\section{Conclusion}
The nonlinear eigenstates $|i\rangle$ that have been defined and discussed in the
present work are time-independent. Since the nonlinear term in
Eq.\ (\ref{eq-Schroe}) is a (square) modulus in the corresponding
time-dependent Schr\"odinger equation, its eigenstates
$\Psi^i\propto u^i\,e^{i\mu_i t/\hbar}$ do still have the standard
time dependence related to the eigenvalue $\mu_i$
of $|i\rangle$. A conceptual problem arises when one considers
the interaction between such states. Indeed, due to their
non-orthogonality defined by Eq.\ (\ref{eq-2Dscalprod}) and illustrated by
Figs.\ \ref{fig1} and \ref{fig5}, one cannot state any more that the system is in
the pure stationary eigenstate $|i\rangle$. It has the probability amplitude $\langle j|i\rangle$
defined by Eq.\ (\ref{eq-theor}) to be also in some other nonlinear state
$|j\rangle$. The statistical weight $w^{ij}=\langle j|i\rangle^2$ then defines the
corresponding mixed state in accordance with Eq.\ (\ref{eq-mixed}).
Actually $w^{ij}$ is quite small: typically less
than  $1\,\% $ (cf.\ Eq.\ (\ref{eq-2Dinterf})).
This might explain the quite acceptable results obtained in GPE systems
when one neglects $w^{ij}$ and assumes that all the boson particles have condensed in the
ground state. Nevertheless, this statement is formally wrong and might yield a
weak measurable departure from the expected GPE ground state particle
density in accurate BEC experiments.

It is tempting  to relate $w^{ij}$
to a mere transition probability. Again, the result is unexpected. Indeed, assume
that the nonlinearity ${\cal N}={\cal N}_i={\cal N}_j$ is small. Therefore each
interaction potential ${\tilde\Phi}^{i,\,j}$ may be regarded as a perturbation
in its corresponding Schr\"odinger equation (\ref{eq-Schroe}). Using standard
time-independent perturbation theory, one obtains ${\Phi}^{i}_{ij}/(E_i-E_j)$
(resp.\ ${\Phi}^{j}_{ji}/(E_j-E_i)$) as the probability amplitude for the
system being in $i$th (resp.\ $j$th) linear eigenstate of energy $E_i$
(resp.\ $E_j$) to be also in the $j$th (resp.\ $i$th) linear eigenstate of
energy $E_j$ (resp.\ $E_i$). One recognizes in the two terms of Eq.\ (\ref{eq-theor})
the simple extrapolation of these probability amplitudes to any value of the
perturbation (i.e.\ of the nonlinearity) while the linear energy
eigenvalues $E_{i,j}$ become the corresponding (nonlinear)
chemical-potential $\mu_{i,j}$. However, due to nonlinear quantum coherence,
there is no simple relationship
between each such probability amplitude and a possible
transition probability. Indeed, for the \{$|i\rangle,|j\rangle$\} two-state system,
exact diagonalization of the perturbed Hamiltonian in the time-dependent
Schr\"odinger equation is possible. It yields in first-order with respect
to $\Phi$ the so-called Fermi golden rule transition probability
$|i\rangle\rightarrow|j\rangle$
\begin{equation}
\label{eq-goldrule}
      {\cal P}_{i \rightarrow j}=
      4 \Biggl[ \frac{\Phi^{i}_{ji}} {E_i-E_j}\Biggr]^2 \sin^2 \frac{(E_i-E_j)t}
      {2\hbar} + o(\Phi^{3})
\end{equation}
or reverse \cite{Cohen88:04}. Since Eq.\ (\ref{eq-goldrule}) is exact, one can average it for
$t\gg \hbar/(E_i-E_j)$ and obtain ${\cal P}_{i \rightarrow j}\sim
2 \Big[ (\Phi^{i}_{ji})/ (E_i-E_j)\Bigr]^2 $ in the lowest order in $\Phi^{2}$;
which, apart from the factor 2, is
indeed the square time-independent probability amplitude that has been previously
obtained.  However, due to Eq.\ (\ref{eq-theor}), the statistical
weight $w^{ij}=\langle j|i\rangle^2$ is \underbar{not} merely equal to the extrapolation of
${\bar {\cal P}}=\frac{1}{2}({\cal P}_{i \rightarrow j}+{\cal P}_{j\rightarrow i})$
to any value of the nonlinearity parameter ${\cal N}$. There remains twice
the probability amplitude
product which lowers ${\bar {\cal P}}$ by more than one order of magnitude.

By focusing our attention on the nonlinear character of two simple cases,
the present paper actually aims to be introductory in the field concerning the reduction
of the original 3N-dimensional ground-state wave vector defining a N-particle linear
quantum system to the 3D one-particle mean-field  nonlinear Schr{\"o}dinger
description based on the mixed ground state given by Eq.\ (\ref{eq-mixed}).
The conclusion that the nonlinear Schr{\"o}dinger equation,
like many other similar ones, is only intended as a recipe
to obtain a solution within an iteration scheme where the problem is considered
linear at each iteration step lies close at hand. However, it remains an open question
whether the direct solution of the nonlinear problem yields properties that are
not obtainable in the iteration scheme, and indeed, if such inherent nonlinear
properties have a physical relevance. While the (mostly numerical)
investigations of the time-dependent GPE mainly concern linear surface excitations of the condensate
about its stationary asymptotic ThF  ground state  profile, true time-dependent nonlinear
structures such as the breathing monopole oscillation mode have been pointed out
\cite{Dalfovo99:463}. These structures are actually solitons,
like in the case of two-BEC time-dependent interferences \cite{Liu00:2294}.
We recall that solitons provide
a paradigm example of a true nonlinear phenomenon that can never be described as a convergent limit of
a linear iterative process. Such seems also the case for the  quantum properties related to
non-orthogonality between the one-particle mean-field nonlinear eigenstates.
Therefore any attempt to investigate the time-dependent
relationship between linear and nonlinear eigenstates of a given quantum system ---
like the recent attempt  to both display particular nonlinear eigenfunctions as global attractors
for all finite-energy solutions and describe quantum transitions between them
\cite{Komech06:09013}--- are welcome. However, only a comparison with experiments
and the solutions of the original linear many-particle problem can shed a
definite light on these questions.

%%%%%%%%%%%%%%%%%%%%%%%%%%%%
\begin{acknowledgments}
      The authors acknowledge financial support from the Icelandic
      Research and Instruments Funds,
      the Research Fund of the University of Iceland and the University of Nice (France).
\end{acknowledgments}
%
%---------------------------------------------
%
%
% \bibliographystyle{apsrev4-1}
\bibliographystyle{apsrev}

\end{document}